\documentclass{optica-article}

\journal{opticajournal} 

\articletype{Research Article}

\usepackage{lineno}

\usepackage{siunitx}
\usepackage{amsmath}

\begin{document}

\title{On the relation between periodically deflected and transverse mode-locked laser beams}

\author{Michael Zwilich,\authormark{1} Jan Wichmann,\authormark{1,*} and Carsten Fallnich}

\address{University of Münster, Institute of Applied Physics, Corrensstr. 2, 48149 Münster, Germany\\
	\authormark{1}These authors contributed equally to this work.}

\email{\authormark{*}jan.wichmann@uni-muenster.de} 


\begin{abstract*}
	Transverse mode-locked (TML) beams exhibit high-speed beam scanning, which motivates a comparison with established beam deflection technologies, such as galvanometer and voice coil scanners. 
	This study explores the hypothesis that TML beams can be regarded as high-speed equivalents of such periodically deflected beams. 
	By analytically modeling the spatiotemporal properties of TML beams as well as experimentally examining periodically deflected beams, both their intensity and phase dynamics were assessed. 
	It is shown that TML beams yield a scanning motion which is shape-invariant upon propagation and have characteristic spatiotemporal phase dynamics.
	While TML-like scanning in intensity can be recreated by combining periodic translational and rotational deflection, beams deflected by mirrors still exhibit different phase dynamics. 
	These findings imply that TML beams cannot simply be categorized as high-speed variants of periodically deflected beams. 
	Nonetheless, TML beams remain applicable to high-speed laser beam scanning, although their phase dynamics have to be considered in phase-sensitive processes.
\end{abstract*}

\section{Introduction}
Scanning laser beams, i.e., laser beams that scan the spatial plane perpendicular to their propagation axis, are commonly used in applications ranging from raster-scanned imaging \cite{Ward2017} and materials processing \cite{Prieto2020} to optical switching \cite{An2021}.
Typically, scanning beams are generated from stationary beams that are then deflected from their initial propagation directions~\cite{Marshall2012}.
In mechanically driven scanning systems incident beams are deflected by mirrors attached to moving elements, such as piezo or galvanometer scanners, reaching spot rates of up to \qty{1}{\MHz}~\cite{Bechtold2013}.
In contrast, solid-state-based deflectors do not employ moving components, but change the beams' propagation directions through electro- or acousto-optical effects, which are induced by externally applied voltages.
These inertia-free electro- or acousto-optic deflectors provide higher spot rates of up to \qty{100}{\MHz}, but are limited in their fields of view~\cite{Bechtold2013}.
In optical phased arrays scanning beams are the result of multiple phase-shifted beamlets interfering in the far field, offering \unit{\MHz} to \unit{\GHz} scanning rates~\cite{Huang2012,Jarrahi2008}.
While all of these deflection systems in and of themselves yield line-scanning beams, the 2D transverse plane can be accessed by using two units with orthogonal deflection directions in series.

Alternatively to the abovementioned deflection techniques, line-scanning beams can also be generated directly in the laser source as so-called transverse mode-locked (TML) beams~\cite{Auston1968a, Auston1968}.
TML beams result from the spatiotemporal interference of higher-order transverse resonator modes, so that scanning beams can be generated in laser systems without moving mechanical components~\cite{Schepers2020, Zwilich2023}. 
This inertia-free design and the already realized spot rates of close to \qty{1}{\GHz} make TML beams candidates for high-speed beam scanning applications~\cite{Schepers2020}.
Multi-\unit{\GHz} scanning rates could even be achieved purely by reducing the resonator length if passive transverse mode-locking was realized, because in TML systems the scanning rate is solely determined by the resonator geometry~\cite{Auston1968a, Siegman1986}. 
One caveat is that TML beams are limited to periodic scanning, in contrast to most mechanical- and solid-state-based beam deflection mechanisms, that also allow for random-access scanning when provided with a non-periodic control signal~\cite{Bechtold2013}.
In that regard, TML systems can be considered more similar to resonant scanning systems, such as resonant microscanners~\cite{Arslan2010,Ataman2006,MasanaoTani2006,Schenk2001} with applications in scanning microscopy~\cite{Ra2007} or medical diagnostics~\cite{Chong2006}.

So far, TML beams have been regarded as laser phenomenons and their generation through mode-locking processes in laser active media has already been discussed~\cite{Smith1974,Agashkov1986,Dukhovnyi1971,Cote1998}.
Here, TML beams are investigated because of the resulting beam scanning instead, which opens up the question of how similar periodically deflected and TML beams are in terms of their spatiotemporal intensity and phase dynamics.
In other words, the hypothesis is addressed that periodically deflected beams can be seen as TML beams -- albeit at lower scanning frequencies.
Besides the fact that both beams exhibit scanning behavior, one might be tempted into making the assumption of them being equivalent based on an analogy.
The field of any stationary displaced Gaussian spot can be expressed as a Poissonian superposition of purely spatial Hermite-Gaussian functions~\cite{Verdeyen1968}.
Therefore, by analogy, it seems logical that any periodically scanning beam, i.e., a beam with temporally varying displacement, can be represented as a Poissonian superposition of Hermite-Gaussian resonator modes, i.e., a TML beam.
Note that, in contrast to the abovementioned purely spatial functions, in resonator modes the spatial Hermite-Gaussian profiles are linked to distinct resonance frequencies~\cite{Siegman1986}, so that their superposition in a TML beam results in a displaced Gaussian spot with a periodic temporal variation of that displacement~\cite{Auston1968a}.
Furthermore, one might hypothesize that the transverse mode frequency shifts found in TML beams are equivalent to the Doppler frequency shifts upon reflection from a moving mirror.
Then, if the mirror was just moved fast enough, the deflected beam would have the same spectrum and hence be indistinguishable from a TML beam with common oscillation frequencies on the order of \unit{\MHz}.

For the following comparison between TML and periodically deflected beams, at first an analytic model of the electric fields of TML beams is used to realistically evaluate their spatiotemporal dynamics in any propagation plane. 
Then, these spatiotemporal intensity and phase dynamics of TML beams are compared to those of periodically deflected beams produced by established scanning systems, namely rotating galvanometer and translating voice coil scanners.
It becomes evident that TML beams are shape-invariant under propagation, whereas periodically deflected beams in general are not. 
And while most deflection systems feature either translational or rotational deflection, the scanning motion of a TML beam is best approximated by a combination of translation and rotation.
Additionally, TML beams exhibit characteristic spatiotemporal phase dynamics, which are not identical to those found in periodically deflected beams.

\section{Spatiotemporal dynamics of TML beams} \label{sec:tml_properties}
The eigenmodes of spherical resonators are well approximated by Hermite-Gaussian $\text{HG}_{m,n}$ modes with their mode indices $m$ and $n$ along the transverse Cartesian coordinates $x$ and $y$, respectively~\cite{Siegman1986}.
While transverse mode-locking generally describes any coherent superposition of transverse resonator modes -- irrespective of the superposed modes and their individual powers -- in the following, a TML beam will be considered to be constituted of $\text{HG}_{m,n=0}\equiv\text{HG}_m$ modes with a Poissonian modal power distribution. 
This representative TML beam was chosen because of its characteristic periodic beam scanning dynamics along one transverse coordinate (here, $x$)~\cite{Auston1968a}.
The spatiotemporal dynamics of such a TML beam have, so far, only been computed analytically at the beam waist, where the wavefront curvature of the constituting modes and propagation effects could be neglected~\cite{Auston1968a, Smith1974, Agashkov1986}. 
Furthermore, the emerging intensity dynamics in the form of beam scanning, resulting from the spatiotemporal interference of the modes, was the main focus of attention in previous studies.
Here, an extended analytic description of TML beams is presented, which is laid out in detail in Sec.~1 of Supplement~1.
It is based on the complete electric field model of Hermite-Gaussian resonator modes including their propagation behavior and both transverse dimensions.
As a result of this approach, TML beam properties can be evaluated at any propagation distance $z$ and not only at the beam waist, i.e., $z=0$.
Furthermore, besides the intensity, the phase was assessed as well, resulting in a complete description of the spatiotemporal dynamics of TML beams.

\begin{figure*}[ht]
	\centering
	\includegraphics{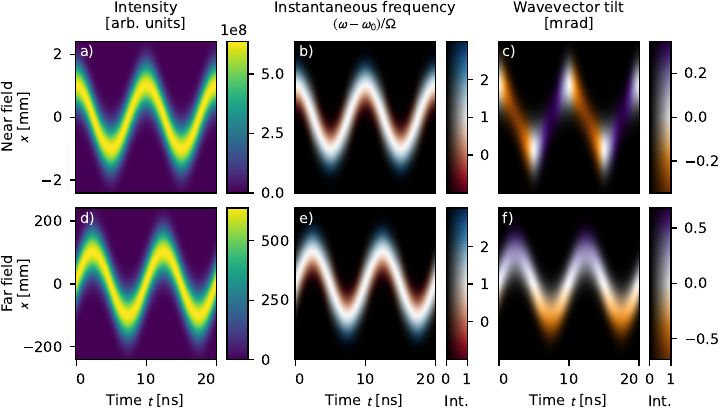}
	\caption{Spatiotemporal dynamics of a TML beam (a-c) in the near field at the beam waist at $z=0$ and (d-f) in the far field at $z=100\,z_\text{R}$. As an example, a TML beam with a transverse mode spacing $\Omega/(2\pi)=\qty{100}{\MHz}$, beam radius $w_0=\qty{1}{\mm}$, fundamental wavelength $\lambda_0=\qty{1064}{\nm}$ (resulting in a Rayleigh length $z_R=\qty{3}{\m}$) and mean mode order $\bar{m}=1$ is shown. In the two rightmost columns the intensity trajectory of the beam ("Int.", normalized from \numrange{0}{1} as the horizontal axes of the colorbars) is encoded as brightness of the pseudocolors, which show the instantaneous frequency and the wavevector tilt, respectively.}
	\label{fig:tml_properties}
\end{figure*}

The amplitude of the electric field of a TML beam consisting of $\text{HG}_m$ modes with a Poissonian modal power distribution and linear modal phases is (see Eq.~(S10) in Supplement~1 for the full analytic expression)
\begin{equation}
	|E_{\text{TML}}(x,z,t)| \propto \exp\left\{-1/2\left[\xi-\xi_0\cos(\Omega t+ \vartheta')\right]^2 \right\}, \label{eq:tml_mag}
\end{equation}
where the transverse coordinate $\xi = \sqrt{2}x/w(z)$ is normalized to the beam radius $w(z)$ of the fundamental mode and $\vartheta'$ summarizes terms not relevant to the discussion at this point.
The most prominent feature of such a TML beam  is spatial scanning, i.e., a Gaussian spot $\exp[-\xi^2]$ that over time $t$ and with a frequency $\Omega$ periodically moves along a straight line with a maximum normalized displacement $\xi_0=\sqrt{2}\Delta x(z)/w(z)$, thus forming a cosine trajectory in the spatiotemporal representation (Fig.~\ref{fig:tml_properties}a and d).
This spatiotemporal oscillation results from the interference of the transverse resonator modes with their distinct spatial profiles $\text{HG}_m$ and associated resonance frequencies $\omega_m=\omega_0+m\Omega$, that are equidistant with the transverse modes spacing $\Omega$.
For simplicity of the discussion presented here, all superposed modes were assumed to have the same mode order $n=0$ in the second transverse dimension, so that along $y$ the emerging TML beam has a Gaussian shape and the explicit $y$-dependence was omitted as it does not affect the periodic scanning motion along $x$.

Evaluating the intensity of a TML beam at different propagation distances $z$ (Fig.~\ref{fig:tml_properties}a and d) reveals that a TML beam can be regarded as shape-invariant in the same way its constituting Hermite-Gaussian modes are~\cite{Borghi2004}.
That is, the intensity distribution remains identical except for a transverse scaling factor, which describes the change in size, and a power factor, which accounts for conservation of energy~\cite{Gori1984, Gori1998}.
In other words, both in the near ($z<z_\text{R}$) and the far field ($z\gg z_\text{R}$), a TML beam exhibits a scanning motion with the same spatial amplitude relative to the fundamental mode size in that plane (note the same visual appearance of Fig.~\ref{fig:tml_properties}a and d).
The temporal shift apparent between near and far field stems from propagation dependent terms in $\vartheta'$ (see Sec.~1 in Supplement~1).
The shape-invariance of TML beams also becomes evident in the normalized oscillation amplitude taking the value $\xi_0 = \sqrt{2}\Delta x(z)/w(z)=\sqrt{2\bar{m}}$, which highlights that the spatial amplitude $\Delta x(z)$ of the scanning motion changes at the same rate as the beam radius $w(z)$, thus forming a constant ratio independent of $z$.

The phase $\phi = \arg(E_\text{TML}) $ of such a TML beam is composed of a term that results from the interference of the transverse modes, as well as terms that describe the phase evolution common to all superposed modes, named "carrier" with frequency $\omega_0$, wavevector $k_0$, Gouy phase shift $\Psi(z)$ and wavefront curvature $R(z)$ (Eq.~(S12) in Supplement~1):
\begin{equation}
	\phi(x,z,t) = \underbrace{\xi_0\sin(\Omega t+\vartheta')\left[\xi-\xi_0\cos(\Omega t+\vartheta')/2\right]\vphantom{\frac{k_0(x^2+y^2)}{2R(z)}}}_\text{TML interference phase} + \underbrace{\omega_0t - k_0z + \Psi(z) - \frac{k_0(x^2+y^2)}{2R(z)}}_\text{carrier phase}. \label{eq:tml_phi}
\end{equation}
The spatiotemporal phase dynamics of a TML beam are best understood by the temporal and spatial derivatives of the beam's phase, which are the instantaneous frequency $\omega = \partial \phi/\partial t$ and the wavevector components $k_{(x,z)} = \partial\phi/\partial (x,z)$, respectively.

Analytic evaluation of the phase $\phi$ reveals that the instantaneous frequency (Eq.~(S13) in Supplement~1) in the center of the scanning spot is always equal to the beam's mean frequency~$\omega_{\bar{m}}$ (Fig.~\ref{fig:tml_properties}b and e, where $\bar{m}=1$). 
With this mean frequency in the center, the instantaneous frequency varies across the transverse cross section at each instant.
Most obvious at the turning points of the cosine trajectory, there is a negative frequency shift on the flank closer to the optical axis and a positive frequency shift on the outside flank.
The magnitude of the transverse frequency variation decreases towards the optical axis, resulting in a Gaussian spot of constant instantaneous frequency at the instants where the spot is at $x=0$.
The magnitude of the frequency shifts relative to the carrier frequency $\omega_0$ across the beam's spot is roughly a few multiples (depending on the power-averaged mean mode order $\bar{m}$) of the scanning frequency $\Omega$, which itself is on the order of tens of \si{\MHz} in common resonator configurations~\cite{Auston1968, Schepers2020, Zwilich2024}.

The wavevector dynamics can be conveniently summarized as $\arctan(k_x/k_z)$, which represents the tilt of the wavevector in the $x$-$z$-plane (Eq.~(S14) and (S15) in Supplement~1).
At the beam waist, a TML beam does not exhibit any wavevector tilt in its turning points (Fig.~\ref{fig:tml_properties}c), i.e., at these instants the wavevector just points in the $z$-direction.
During the scanning motion the wavevector is tilted positively and negatively in an alternating fashion with the greatest tilt magnitude on the optical axis ($x=0$).
The existence and evolution of the wavevector tilt in the near field has been shown to be of paramount importance when coupling TML beams to single-mode fibers because of their wavefront-selective mode-matching~\cite{Wichmann2024}.
In the far field the wavevector tilt evolves differently: it changes along $x$ from negative tilts in the lower turning points to positive tilts in the upper turning points (Fig.~\ref{fig:tml_properties}f).
This far field wavevector distribution is plausible when viewing the TML beam as a superposition of plane waves with propagation directions given by the wavevector tilts in the near field: the positively and negatively tilted wavefronts propagate away from the optical axis ending up at the upper and lower turning points, respectively, while the non-tilted wavefronts propagate parallel to the optical axis and thus remain close to it.

To summarize, the three specific features characterizing a TML beam constituted by $\text{HG}_m$ modes are (1) periodic beam scanning along a straight line in the transverse plane, (2) shape-invariance upon propagation, and (3) spatiotemporal phase dynamics, apparent in variations of instantaneous frequency as well as wavevector tilt.

\section{Spatiotemporal dynamics of periodically deflected beams}
\subsection{Experimental setup} \label{subsec:exp_setup}
A periodic displacement of a laser spot, as observed with TML beams (Fig.~\ref{fig:tml_properties}a), is also achieved by periodically deflecting a beam from a moving mirror, such as a galvanometer scanner.
Therefore, one might wonder, if TML beams can simply be regarded as high-speed variants of such routinely generated periodically deflected beams and if both types of scanning beams have equivalent electric field distributions.
In the following, this point is addressed by comparing the spatiotemporal properties of TML beams discussed in Sec.~\ref{sec:tml_properties} with those of beams periodically deflected from translated and rotated mirrors.
Because the authors are not aware of an analytic framework that adequately describes the electric field dynamics of so deflected beams and initial numeric calculations turned out to be too computationally expensive, an experimental approach was chosen to investigate the spatiotemporal dynamics of periodically deflected beams.

\begin{figure}[ht]
	\centering
	\includegraphics{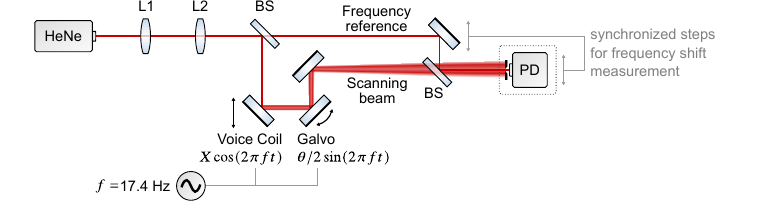}
	\caption{Schematic of the scanning beam experimental setup. Scanning beams are generated by periodic deflection from a translational (voice coil) and a rotational (galvo) mirror. Spatiotemporal intensity distributions are captured by a spatial photodetector (PD) scan. The instantaneous frequency is measured from the beating with a reference beam. L: lens, BS: 50:50 beamsplitter.}
	\label{fig:setup}
\end{figure}

Scanning beams were generated by deflecting the beam of a Helium-Neon laser source (HeNe, wavelength $\lambda\cong\qty{633}{\nm}$) from a combination of two mirrors: one attached to a voice coil scanner, the other to a galvanometer scanner, allowing for periodic deflection by translational and rotational motion, respectively (see Fig.~\ref{fig:setup} for the experimental setup).
These mirror-based deflection systems were chosen for their ease of use and to evaluate the effects of the fundamental motions of translation as well as rotation, which are also exhibited by other deflection systems~\cite{Marshall2012}.
Both deflection mechanisms were operated at a frequency $f=\qty{17.4}{\Hz}$, which was determined by the resonance of the voice coil scanner. 
The translational mirror was moved with a displacement amplitude $X$ according to $X\cos(2\pi ft)$, which also describes the periodic displacement of the reflected beam.
The rotational mirror performed the angular motion $\theta/2\sin(2\pi ft)$, so that the reflected beam was displaced by $\theta z \sin(2\pi ft)$, using the small angle approximation $\tan\theta \approx \theta$ as the applied rotation amplitudes $\theta$ were on the order of \si{\milli\radian} only.
This two-mirror-setup allowed to investigate the effects of their combined motion as well as the effects of translation and rotation separately by setting the respective other motion amplitude to zero.
The deflection mirrors were placed in close proximity to the beam waist, which was formed after passing through the telescope composed of lenses L1 and L2.
The beam spot radius $w_0=\qty{0.3}{\mm}$ with its Rayleigh length $z_\text{R}=\qty{0.44}{\m}$ was chosen to allow for practical observations in the near field as well as in the far field.

The spatiotemporal intensity distribution of a generated periodically deflected beam was measured by stepping a photodetector along the beam's scanning axis $x$ and recording a temporal trace with an oscilloscope at every step (all the while blocking the reference beam introduced below).
The intensity measurements acquired at different spatial positions were synchronized in time by using the \qty{17.4}{\Hz} signal driving the voice coil scanner as a trigger.
To gain insight into the phase dynamics, the periodically deflected beam was interfered with a frequency reference beam, which was split off from the same laser source~\cite{Drain1980, Kroschel2020}.
The resulting beating signal changed periodically over time, because the deflected beam experienced a Doppler frequency shift on reflection from the moving mirrors~\cite{Sommerfeld1954, Bernal2007}, making the beam's instantaneous frequency the natural choice for an easy to measure phase-related quantity.
Thus, while simultaneously stepping the photodetector and the reference beam transversely along the scanning beam, at each spatial position the instantaneous frequency was determined from the periodic beating signal via a short-time Fourier transform.
The spatiotemporal intensity and frequency distributions were measured in the near field ($z=0.5\,z_\text{R}$) as close as possible to the beam waist (defining $z=0$) and in the far field ($z=10\,z_\text{R}$) to evaluate the influence of propagation.

\subsection{Scanning behavior and shape-invariance of periodically deflected beams} \label{subsec:exp_intensity}
\begin{figure}[ht]
	\centering
	\includegraphics{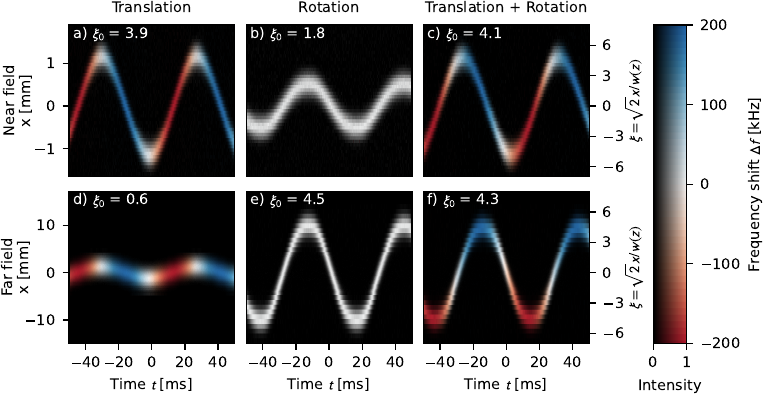}
	\caption{Scanning motion of beams deflected by (a,d) a periodically translated mirror with $X=\qty{1.01}{\mm}$, (b,e) a rotated mirror with $\theta=\qty{2.4}{\milli\radian}$ or (c,f) a translation-rotation mirror pair with the same parameters. The (a-c) near and (d-f) far field measurements were carried out at $z=0.5\,z_\text{R}$ and $z=10\,z_\text{R}$, respectively. The Doppler frequency shift $\Delta f$ due to the reflection from a moving surface is given in pseudocolors while the intensity (normalized from \numrange{0}{1} as the horizontal axis of the colorbar) is encoded as brightness.}
	\label{fig:galvocoil}
\end{figure}

As expected, periodic translation and rotation of the deflection mirrors each resulted in transverse beam scanning along a straight line in any plane, forming a cosine trajectory in the spatiotemporal representation.
However, the scanning trajectories were not shape-invariant, i.e., the normalized oscillation amplitudes $\xi_0(z)=\sqrt{2}\Delta x(z)/w(z)$ changed upon propagation (compare Fig.~\ref{fig:galvocoil}a with d, and b with e).
When only deflecting the beam with translational mirror motion, the normalized spatial oscillation amplitude $\xi_0(z)$ decreased upon propagation from \num{3.9} to \num{0.6} because the displacement $\Delta x = X$ was constant, whereas the beam radius $w(z)$ increased.
Conversely, for rotational deflection, $\Delta x(z)=\theta z$ does scale with $z$, and due to that very dependence on $z$ the normalized oscillation amplitude increased from \num{1.8} in the near field to \num{4.5} in the far field.
Exactly at the beam waist, the oscillation amplitude would even be zero, i.e., $\xi_0(z=0)=0$, which again highlights that the scanning pattern of a beam displaced through rotation also changes with propagation.
This is unlike TML beams, where the normalized spatial oscillation amplitude $\xi_0$ is constant, because both the spot size and the displacement change at the same rate.

A TML-like, shape-invariant scanning behavior was achieved by combining translation (for near-field displacement) and rotation (for far-field divergence) of the two deflection mirrors with $\pi/2$ phase difference (Fig.~\ref{fig:galvocoil}c and f).
Additionally, see Sec.~2 in Supplement~1 for an analytic derivation and experimental verification.
The oscillation amplitude $\xi_0$ relative to the diverging beam size remained constant from the near to the far field, if the condition $X/w_0=\theta/\theta_0$ was fulfilled, i.e., if the ratio between displacement $X$ and spot radius $w_0$ at the beam waist was equal to the ratio between beam deflection angle $\theta$ and inherent beam divergence $\theta_0=\lambda/(\pi w_0)$.
Note the same visual appearance of the scanning motion in the near field and the far field (compare Fig.~\ref{fig:galvocoil}c and f), which also becomes apparent in the approximately constant normalized oscillation amplitude of $\xi_0=\num{4.1}\approx\num{4.3}$. 
This shape-invariant beam scanning was similar to the spatiotemporal intensity dynamics seen in TML beams (Fig.~\ref{fig:tml_properties}a and d) -- briefly putting aside their different scanning frequencies (here, \qty{17.4}{\Hz} and \qty{100}{\MHz}).
This similarity provides the insight that the intensity dynamics of TML beams, i.e., a transverse scanning motion which is shape-invariant upon propagation, can be understood as a Gaussian beam that is both translated and rotated periodically at its waist. 

To summarize this section, combined translational and rotational deflection produces line-scanning beams with shape-invariant intensity distributions upon propagation, which could therefore be seen as TML beams with correspondingly chosen parameters.
Hence, when only considering their intensity dynamics, beams periodically deflected through translation and rotation can be regarded as equivalent to TML beams in any plane.
Note that, these findings follow directly from the type of deflection motion, i.e., translation and rotation, and therefore not only apply to the representative choices of voice coil and galvanometer scanner but to other deflection mechanisms as well.
The commonly used acousto- or electro-optic deflectors or optical phased arrays, for example, only provide angular deflection.
Therefore, the beams deflected by these solid-state based deflectors alone do not replicate the shape-invariant scanning of TML beams, because it requires not only rotational but also translational beam motion.

\subsection{Phase dynamics of periodically deflected beams} \label{subsec:exp_phase}
As pointed out in Sec.~\ref{subsec:exp_setup}, instead of directly comparing beams based on their electric fields, here the easily measurable instantaneous frequency, i.e., the temporal derivative of the phase, is used as a proxy.

The periodically deflected beams exhibited temporally dependent frequency shifts as a result of the Doppler effect on reflection from a moving surface~\cite{Sommerfeld1954}.
The beam deflected by the translational mirror showed a frequency shift that evolved over time between zero at the turning points, where the mirror was at rest, and $\pm\qty{195}{\kHz}$ at the scanning midpoints, where the mirror velocity was at its maximum (Fig.~\ref{fig:galvocoil}a and d).
The sign of the frequency shift depended on the direction of motion, which was inferred from the change in beam position, resulting in an increased (decreased) frequency when the mirror moved towards (away from) the beam.
At each instant, the optical frequency was constant along the beam's cross section, because the mirror velocity -- and hence the introduced Doppler shift -- is independent of $x$ in translational motion.

In contrast, the beam deflected through rotation was expected to show a transverse variation in Doppler shift and instantaneous frequency, because in rotational motion the mirror velocity depends on the transverse position relative to the axis of rotation. 
However, the beam deflected through rotation (Fig.~\ref{fig:galvocoil}b and e) did not show any measurable frequency shift, because the mirror velocity of $<\qty{0.1}{\mm/s}$ across the beam was low compared to $\qty{19.5}{\mm/s}$ for the translational motion. 
Hence, the expected Doppler shift was also comparatively low with $<\qty{0.1}{\kHz}$, which could not be resolved experimentally, because the signal generated by the scanning beam passing the static photodetector was too short in temporal duration (about \qty{3}{\ms}) to obtain sufficiently high frequency resolution.
Furthermore, even though the instantaneous frequency distribution of a beam deflected through rotation in principle exhibits a transverse frequency shift, it still differs from that of a TML beam: taking as an example the instant when the mirror is in its maximum angular orientation, so that the beam is maximally deflected and forms a turning point of the scanning motion.
At this instant, the mirror's velocity is zero and so the deflected beam experiences no Doppler shift, which leads to a maximally displaced spot in the far field without any frequency shift -- in stark contrast to TML beams, where the transverse frequency shift is most prominent in the turning points (Fig.~\ref{fig:tml_properties}b and e).
Therefore, even if not resolved in the measurements, the instantaneous frequency distributions of beams deflected through rotation are unlike that of TML beams.

Due to the low Doppler shift from rotation, the shape-invariant scanning beam produced via combined translation and rotation (Fig.~\ref{fig:galvocoil}c and f) exhibited in principle the same frequency dynamics in the near field as the translated beam, i.e., zero frequency shifts at the turning points and maximum shifts in-between.
In the far field, however, the locations of the frequency shifts with respect to the scanning motion were reversed, and maximum frequency shifts were measured at the turning points.
This behavior could be observed because the Doppler shift of the beam at any point in time mainly depended on the velocity of the translational deflection mirror, whereas the position of the beam in the far field mainly depended on the angular position of the rotational deflection mirror.
To give an example, at $t=0$ the translational mirror was maximally displaced, but momentarily at rest, thus introducing no frequency shift.
At the same instant, because of the $\pi/2$ phase difference between translation and rotation, the rotational mirror was in its central \ang{45} position, consequently, this beam segment without a frequency shift propagated parallel to the optical axis into the far field.
Therefore, moments of zero frequency shifts would occur at the lower and upper turning points in the near field, but at $x=0$ in the limit of the infinitely distant far field, where the initial translational displacement can be neglected.
In the experiment slight shifts of these locations were observed, because the near field was evaluated at $0.5\,z_\text{R}$ instead of $z=0$ in order to place the optics necessary for the beating measurement and the far field was evaluated at $10\,z_\text{R}$ instead of infinity for practical reasons.

Before comparing the instantaneous frequency distributions of periodically deflected beams to those of TML beams, the properties of the latter are briefly reiterated for convenience.
The instantaneous frequency distribution of a TML beam (i) exhibits frequency shifts along its transverse beam cross section (except for $x=0$), (ii) shows frequency shifts on the order of magnitude of the scanning frequency $\Omega$ and (iii) is shape-invariant upon propagation.
By comparison, in the presented measurements the instantaneous frequency distributions of periodically deflected beams differed in all three of these aspects: (i) there was no measurable frequency shift along the beam's transverse cross section, i.e., at each point in time the instantaneous frequency was the same and hence independent of the transverse coordinate $x$, (ii) the absolute value of the observed frequency shifts (\qty{195}{\kHz}) was several orders of magnitude greater than the scanning frequency (\qty{17.4}{\Hz}) and (iii) the spatiotemporal frequency distribution changed with propagation distance (compare Fig.~\ref{fig:galvocoil}c and f).
This means, even in the cases where the spatiotemporal intensity distributions of periodically deflected and TML beams were similar to each other, their instantaneous frequency dynamics and therefore also their phase dynamics exhibited qualitative differences.
These fundamental disparities would exist just as well, if the deflection took place at \unit{\MHz} scanning rates similar to TML beams or for (unrealistically low) TML scanning rates on the order of \unit{\Hz}.

Initially, the hypothesis was proposed that any line-scanning beam can be regarded as a TML beam because of their similar intensity scanning dynamics.
But the presented results led to the conclusion that, in regard to their electric fields, periodically deflected beams in general cannot be viewed as TML beams and vice versa.
Furthermore, the proposition that the transverse mode frequency shifts in TML beams correspond to Doppler frequency shifts in deflected beams, has to be rejected, because the instantaneous frequency of beams deflected from moving mirrors were found to be qualitatively different from that of a TML beam.
While this does not prove that the spatiotemporal field dynamics of TML beams are unique or that they could not be achieved by means of (a combination of) different deflection approaches, the unequal phase dynamics clearly demonstrate that not every scanning beam is a TML beam.

\section{Conclusion}
Transverse mode-locked (TML) beams are best known for generating moving laser spots, which, at first glance, resemble line-scanning beams produced by established deflection techniques, such as galvanometer scanners.
Therefore, this study was motivated by the naturally arising question, if TML beams can be seen as high-speed variants of periodically deflected beams or, vice versa, line-scanning beams can be expanded into a modal basis and be regarded as TML beams.
Additionally, the potential of TML beams to achieve multi-\unit{\GHz} scanning rates further inspired exploration of their spatiotemporal dynamics, as their phase and propagation properties had not yet been examined.

First, the electric fields of TML beams were modeled analytically, revealing not only their shape-invariant propagation but also phase dynamics, such as spatiotemporal variations in instantaneous frequency and oscillations of the wavevector direction, both in sync with the spatial beam motion.
Second, for comparison with TML beams, scanning beams were generated experimentally in a representative deflection system consisting of a translating voice coil and a rotating galvanometer scanner.
The shape-invariant scanning behavior of TML beams was found to be replicated by combined translational and rotational deflection, where it was crucial to match the deflection parameters to the beam's inherent size and divergence, respectively.
Despite similar intensity dynamics, the phase dynamics of the periodically deflected beams were still unlike that of TML beams, which makes the beams an example of the ambiguity between intensity and electric field: TML and periodically deflected beams are represented by qualitatively different fields that just happen to result in similar intensities -- even when accounting for their respective scanning frequencies. 
In conclusion, the characteristic spatiotemporal fields of TML beams could not be replicated with translational and rotational mirror-based deflection, which highlights that TML beams in general are not high-speed variants of periodically deflected beams when considering their complete field dynamics including their phases.

These differences in the spatiotemporal phase distributions should be carefully considered for phase-sensitive measurements, such as phase contrast microscopy~\cite{Jungerman1984,See1985}, when using TML beams instead of established laser-scanning techniques.
But as most scanning beam applications are intensity-based~\cite{Marshall2012}, TML remains a potential technique for high-speed laser beam scanning with already demonstrated spot rates close to \qty{1}{\GHz}~\cite{Schepers2020}.
To increase applicability of TML beams, overcoming the current limitation of scanning along only one axis as well as achieving greater deflection amplitudes -- and therefore wider fields of view -- are key objectives for further development of TML systems.

\begin{backmatter}
	\bmsection{Funding} Open Access Publication Fund of the University of Münster.
	\bmsection{Acknowledgments} We acknowledge support from the Open Access Publication Fund of the University of Münster.
	\bmsection{Disclosures} The authors declare no conflicts of interest.
	\bmsection{Data availability} Data underlying the results presented in this paper are not publicly available at this time but may be obtained from the authors upon reasonable request.
	\bmsection{Supplemental document} See Supplement~1 for supporting content.
\end{backmatter}

\bibliography{bib}	
\end{document}


\maketitle
\section{Full analytic description of TML beams}\label{sec:tml_analytic}

\subsection{Hermite-Gaussian resonator modes}
Before exploring the field dynamics of transverse mode-locked (TML) beams, the notation for the electric fields of the constituting Hermite-Gaussian resonator modes is presented.
Throughout this section, $x$ and $y$ are transverse spatial coordinates, $z$ is the longitudinal coordinate along which propagation takes place, and $t$ is time.
For the following derivations it is helpful to write the electric field of a Hermite-Gaussian resonator mode of order $m,n$ in the form (adapted from Ch.~17, Eq.~(41) in~\cite{Siegman1986})
\begin{equation}
	E_{m,n}(x,y,z,t) = u_m(x,z,t) u_n(y,z,t) \exp[i\varphi] \label{eq:E_mn}.
\end{equation}
Here, the mode-order dependent transverse spatial factors $u_m$ and $u_n$ are separated from the mode-order independent phase $\varphi = \left[\omega_0t-k_0z+\Psi(z)-k_0\left(x^2+y^2\right)/(2R(z))\right]$, where $\omega_0$ is the frequency, $k_0$ the wavevector magnitude, $\Psi(z)$ the Gouy phase shift and $R(z)$ the wavefront curvature.  
These quantities, as well as the beam radius $w(z)$ of the fundamental mode $E_{0,0}$, scale upon propagation relative to the radius $w_0$ at the beam waist and the resulting Rayleigh length $z_\text{R}=\pi w_0^2/\lambda$:
\begin{equation}
	w(z) = w_0\sqrt{1+\left(\frac{z}{z_\text{R}}\right)^2},\quad 
	R(z) = z\left[1+\left(\frac{z_\text{R}}{z}\right)^2\right],\quad 
	\Psi(z) = \arctan\left(\frac{z}{z_\text{R}}\right).
\end{equation}
The mode-order dependent Hermite-Gaussian amplitude and phase terms are accounted for by (analogous for $u_n(y,z,t)$)\footnote{The explicit dependencies of the argument $\vartheta(x,z,t)$ and $N_m(z)$ are omitted for ease of readability.}
\begin{align}
	u_m(x,z,t) &= N_m H_m(\xi) \exp[-\xi^2/2] \exp[im\vartheta]\quad\text{ with} \label{eq:um}\\ 
	N_m(z) &= \sqrt{\sqrt{\frac{2}{\pi}}\frac{1}{2^mm!w(z)}}\quad\text{and}\quad \vartheta(x,z,t) = \left[\Omega t-Kz+\Psi(z)-Kx^2/(2R(z))\right]. \nonumber
\end{align}
$H_m$ is the $m$-th order Hermite polynomial and $\xi = \sqrt{2}x/w(z)$ is the $x$-coordinate normalized by the fundamental mode radius $w(z)$.
The term $N_m$ is chosen, such that the power across the transverse plane is normalized to 1, i.e., $\iint |E_{m,n}|^2\,\mathrm{d}x\,\mathrm{d}y = 1$.
The resonance frequencies $\omega_m = \omega_0 + m\Omega$ of transverse resonator modes are distributed linearly with a spacing of $\Omega$ and so are the magnitudes of their wavevectors $k_m = \omega_m/c = k_0 + mK$ with $K=\Omega/c$ and the speed of light $c$. 

\subsection{TML beams as Hermite-Gaussian mode superpositions}
Here, TML beams will be considered to be the coherent superposition of Hermite-Gaussian resonator modes $E_{m,n}$ (with $n$ being constant).
In earlier derivations of the spatiotemporal dynamics of TML beams, simplified expressions were used for the electric fields of the individual modes, with $E_{m,0}(\xi, t, z=0)\propto H_m(\xi)\exp[-\xi^2/2]\exp[i\omega_m t]$~\cite{Auston1968a, Smith1974, Agashkov1986}.
These were restricted to a single transverse dimension $\xi$ for simplicity and  -- more importantly -- to the beam waist $z=0$.
Thus, propagation effects were neglected.
Here, to overcome these limitations, the complete electric field descriptions of Hermite-Gaussian resonator modes (see Eq.~(\ref{eq:E_mn})) were taken as a basis, such that the electric field of a TML beam constituted by $E_{m,n}$ modes can be expressed as
\begin{equation}
	E_{\text{TML}}(x,y,z,t) = \sum_{m=0}^\infty a_m E_{m,n} = u_n \exp[i\varphi] \sum_{m=0}^\infty a_m N_m H_m(\xi) \exp[-\xi^2/2] \exp[im\vartheta],
	\label{eq:e_tml_sum}
\end{equation}
where each mode $E_{m,n}$ is weighted by an amplitude $a_m$. When assuming a Poissonian modal power distribution $|a_m|^2 = \exp(-\bar{m})\bar{m}^m/m!$ of mean mode order $\bar{m}$, one can rewrite
\begin{equation}
	a_m N_m = N_0 \exp\left[-\xi_0^2/4\right] \frac{\xi_0^m}{2^m m!},
\end{equation}
while defining $\xi_0 = \sqrt{2\bar{m}}$, such that 
\begin{equation}
	E_{\text{TML}}(x,y,z,t) = u_n \exp[i\varphi] N_0 \exp\left[-\xi^2/2\right] \exp\left[-\xi_0^2/4\right] \sum_{m=0}^{\infty} \text{H}_m(\xi) \exp\left[im\vartheta\right] \frac{\xi_0^m}{2^m m!}. 
\end{equation}
After identifying $b=\xi_0\exp[i\vartheta]/2$, the exponential generating function of Hermite polynomials 
\begin{equation}
	\sum_{m=0}^{\infty}\text{H}_m(\xi)\frac{b^m}{m!} = \exp\left[2\xi b-b^2\right]
\end{equation}
can be applied, such that the sum over $m$ is resolved. Then the complex exponentials $\exp[i\vartheta]$ and $\exp[i2\vartheta]$ are expanded into sine and cosine. Using the double-angle identities $\cos(2\vartheta)=2\cos^2(\vartheta)-1$ and $\sin(2\vartheta)=2\sin(\vartheta)\cos(\vartheta)$ allows to refactor $-\xi/2+\xi\xi_0\cos(\vartheta)-\xi_0\cos^2(\vartheta)/2\allowbreak= -1/2[\xi-\xi_0\cos(\vartheta)]^2$.
After some final algebra, the electric field of a TML beam reads:
\begin{equation}
	E_{\text{TML}}(x,y,z,t) = u_n \exp[i\varphi] N_0 \exp\left\{-1/2\left[\xi-\xi_0\cos(\vartheta)\right]^2 + i\xi_0\sin(\vartheta) \left[\xi - \xi_0\cos(\vartheta)/2\right]\right\}.
	\label{eq:tml}
\end{equation}

\subsection{Amplitude and phase dynamics of TML beams}
While the amplitude dynamics, i.e., the beam scanning property of TML beams, have been investigated in earlier works, the phase dynamics have not yet been fully explored.
Here, we present the analytic description of a TML beam's phase by means of its spatiotemporal derivatives, i.e., instantaneous frequency and spatial wavevector components.
For this evaluation of amplitude and phase dynamics, the electric field of a TML beam is written as
\begin{equation}
	E_\text{TML}(x,y,z,t) = |E_\text{TML}| \exp[i\phi],
\end{equation}
with the amplitude
\begin{equation}
	|E_\text{TML}(x,y,z,t)| = \underbrace{N_0 \exp\left\{-\frac{1}{2}\left[\xi-\xi_0\cos(\vartheta)\right]^2 \right\}\vphantom{\left(\frac{\sqrt{2}y}{w(z)}\right)}}_{\text{Beam scanning in $x$-direction}} \underbrace{N_n H_n\left(\frac{\sqrt{2}y}{w(z)}\right)\exp\left[\frac{-y^2}{w^2(z)}\right]}_{\text{Single mode of order $n$ in $y$-direction}}.
	\label{eq:tml_amplitude}
\end{equation}
The interference of the Hermite-Gaussian resonator modes with their distinct spatial profiles and resonance frequencies yields periodic beam scanning (see Fig.~1a and d in the main document) at the frequency $\Omega$, that is contained in the argument $\vartheta$ (see Eq.~(\ref{eq:um})). 
To highlight this temporal dependence, in the main document the argument of the sine and cosine functions is given as
\begin{equation}
	\Omega t + \vartheta' = \vartheta.
\end{equation}

Correspondingly, the phase of a TML beam can be read off from Eq.~(\ref{eq:tml}) to be
\begin{align}
	&\phi(x,y,z,t) = \nonumber\\ &\underbrace{\xi_0\sin(\vartheta)\left[\xi-\xi_0\cos(\vartheta)/2\right]\vphantom{\left[\frac{Ky^2}{2R(z)}\right]}}_{\text{TML phase}}
	+ \underbrace{\omega_0t - k_0z + \Psi(z) - k_0\frac{x^2+y^2}{2R(z)}}_{\text{Carrier phase}}
	+ \underbrace{n\left[\Omega t-Kz+\Psi(z)-\frac{Ky^2}{2R(z)}\right]}_{\text{Modal phase of order $n$}}.
	\label{eq:tml_phase}
\end{align}
The instantaneous frequency $\omega$ is the temporal derivative of the beam's phase.
\begin{equation}
	\omega(x,z,t) = \partial\phi/\partial t = \omega_0 + n\Omega + \bar{m}\Omega\sin^2(\vartheta) + \xi_0\Omega \left[\xi - \xi_0\cos(\vartheta)/2\right]\cos(\vartheta) \label{eq:omega}
\end{equation}
Its spatiotemporal distribution in the near and far field is shown in Fig.~1b and e in the main document.
Note that, the argument $\vartheta$ defined in Eq.~(\ref{eq:um}) depends on time, which via chain rule for derivation leads to the modal interference terms in Eq.~(\ref{eq:omega}), in addition to the carrier frequency $\omega_0$ and the frequency shift $n\Omega$ based on transverse mode order in $y$-direction.

The $x$- and $z$-components of the wavevector $k$ are the spatial derivatives of $\phi$:
\begin{equation}
	\begin{split}
		k_x(x,t) = \partial\phi/\partial x = \{&-2\xi_0Kx\left[\xi-\xi_0\cos(\vartheta)/2\right]\cos(\vartheta)\\
		&+\xi_0\left[2R\xi/x-\xi_0Kx\sin(\vartheta)\right]\sin(\vartheta)-2k_0x\}/(2R),
	\end{split}
 \label{eq:kx}
\end{equation}
\begin{equation}
	\begin{split}
	k_z(x,y,z,t) &= \partial\phi_\text{TML}/\partial z = - k_0 - \frac{k_0(x^2+y^2)}{2R^2}(R/z-2) + \frac{2}{k_0w^2} \\
	&-\xi_0\left[\xi-\xi_0\cos(\vartheta)/2\right]\left[\frac{Kx^2}{2R^2}(R/z-2)+K-\frac{2}{k_0w^2}\right]\cos(\vartheta) \\
	&- \xi_0\left\{\frac{4\xi z}{(k_0ww_0)^2}+ \xi_0/2\left[\frac{Kx^2}{2R^2}(R/z-2)+K-\frac{2}{k_0w^2}\right]\sin(\vartheta)\right\}\sin(\vartheta) \\
	&- n\left[\frac{Ky^2}{2R^2}(R/z-2)+K-\frac{2}{k_0w^2}\right]. 
\end{split} \label{eq:kz}
\end{equation}
Again, most terms arise from the repeated application of the chain rule, because $\vartheta$, $\Psi$ and $R$ are dependent on $x$ and/or $z$.\footnote{The $z$-dependence of the spot radius $w(z)$ and the wavefront curvature $R(z)$ was omitted in Eqs.~(\ref{eq:kx}), (\ref{eq:kz}) and (\ref{eq:ky}) for better readability.}
Eqs.~(\ref{eq:kx}) and (\ref{eq:kz}) are conveniently summarized as $\arctan(k_x/k_z)$ representing the wavevector's tilt angle in the $x$-$z$-plane, revealing that the wavevector of a TML beam oscillates in direction (see Fig.~1c and~f in the main document).
Further discussion of the spatiotemporal dynamics in TML beams can be found in Sec.~2 of the main document.
For completeness, the $y$-component of the wavevector shows no spatiotemporal dynamics, because all superposed modes in the TML beam defined by Eq.~(\ref{eq:e_tml_sum}) have the same mode order $n$ in $y$-direction:
\begin{equation}
	k_y(y) = \partial\phi/\partial y = -\frac{y(k_0+nK)}{2R}.
	\label{eq:ky}
\end{equation}

\cleardoublepage
\section{Shape-invariance of periodically deflected beams}
While periodically deflected beams have a myriad of applications, their electric field or even intensity dynamics are only rarely analytically considered.
Furthermore, the authors are not aware of any expression, which describes the intensity of beam scanning resulting from combined translational and rotational deflection.
Here, for analytic support of the presented experiments, a Gaussian beam deflected by a translated and rotated mirror is considered.
It is shown that this combination of different types of deflection can lead to a scanning motion which is shape-invariant upon propagation.

\begin{figure}[h]
	\centering
	\includegraphics[scale=1.1]{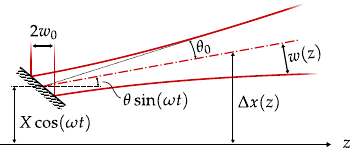}
	\captionof{figure}{Reflection of a Gaussian beam from a translated and rotated mirror.}
	\label{fig:reflection}
\end{figure}

The displacement of a Gaussian beam that is periodically deflected at a frequency $\omega=2\pi f$ through translation and rotation, both $\pi/2$ out of phase relative to each other, is
\begin{equation}
	\Delta x(z) = X\cos(\omega t) + \theta z\sin(\omega t),
	\label{eq:dx}
\end{equation}
where $X$ is the translational displacement and $\theta z$ is the displacement caused by rotation after propagation along $z$.
Fig.~\ref{fig:reflection} provides a visualization of the geometry.

The harmonic addition theorem $a\cos(x)+b\sin(x)=\operatorname{sgn}(a)\sqrt{a^2+b^2}\allowbreak\cos(x+\arctan(-b/a))$ allows to rewrite Eq.~(\ref{eq:dx}) to
%
\begin{equation}
	\Delta x(z) = \sqrt{X^2+\theta^2 z^2} \cos(\omega t + \arctan(-\theta z/X)),
\end{equation}
where, without loss of generality, $X>0$ was assumed for simplicity.
The spot radius $w(z)$ of the underlying, deflected Gaussian beam changes upon propagation as well:
\begin{equation}
	w(z) = w_0\sqrt{1+\left(\frac{z}{z_R}\right)^2}, \quad \text{with } z_R = \frac{\pi w_0^2}{\lambda}.
\end{equation}
This can be rewritten using the beam's divergence $\theta_0=\lambda/(\pi w_0)$, such that $z_R=w_0/\theta_0$ and therefore
\begin{equation}
	w(z) = \sqrt{w_0^2 + \theta_0^2 z^2}.
\end{equation}
The scanning beam shall remain shape-invariant, i.e., its displacement $\Delta x(z)$ relative to the spot radius $w(z)$ shall be independent of z and hence a constant, which will be denoted as $\sqrt{\bar{m}}$:\footnote{The proportionality constant was named $\sqrt{\bar{m}}$ to highlight a connection to TML beams: a TML beam with Poissonian modal powers and mean mode order $\bar{m}$ also exhibits periodic transverse beam scanning, that is shape-invariant upon propagation with a normalized amplitude $\Delta x(z)/w(z)=\sqrt{\bar{m}}$ (see Sec.~1: $\xi_0=\sqrt{2\bar{m}}=\sqrt{2}\Delta x(z)/w(z)$).}
\begin{equation}
	\frac{\Delta x(z)}{w(z)} = \sqrt{\bar{m}}.
\end{equation}
The temporal dependence $\cos(\cdot)$ can be ignored when evaluating the spatial oscillation amplitude along propagation:
\begin{align}
	\frac{\Delta x(z)}{w(z)} = \frac{\sqrt{X^2+\theta^2 z^2}}{\sqrt{w_0^2 + \theta_0^2 z^2}} \Rightarrow \frac{X^2+\theta^2 z^2}{w_0^2 + \theta_0^2 z^2} = \bar{m}.
	\label{eq:dx_w_ratio}
\end{align}
Separately comparing the terms in these second-order polynomials of $z$ leads to the condition for shape-invariance:
\begin{equation}
	\frac{X}{w_0} = \frac{\theta}{\theta_0} = \sqrt{\bar{m}}.
	\label{eq:shapeinvariance}
\end{equation}

\begin{figure}[ht]
	\centering
	\includegraphics{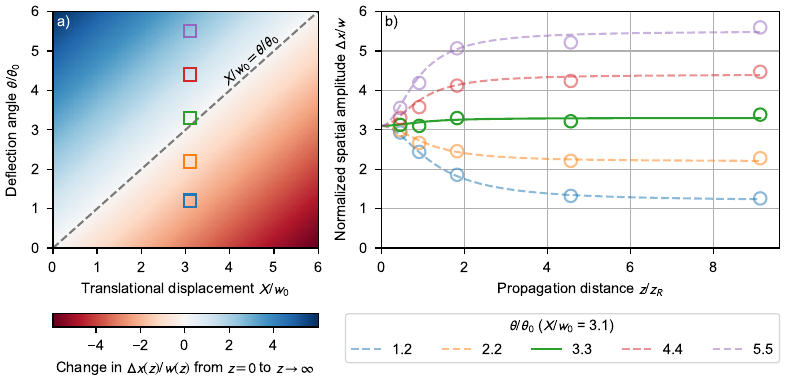}
	\caption{\textbf{a)} Calculation based on Eq.~(\ref{eq:dx_w_ratio}). Scanning beams are shape-invariant, whenever the condition $X/w_0=\theta/\theta_0$ is fulfilled. The parameters of the measurements shown in b) are denoted by colored squares. \textbf{b)} Measured spatial oscillation amplitudes of deflected beams. Shape-invariant scanning is best approximated in the case of the highlighted green curve, because here $X/w_0=3.1\approx\theta/\theta_0=3.3$, leading only to a slight increase of $\Delta x/w$ from the near field to the far field. Circles denote measurements, the corresponding lines are plotted according to Eq.~(\ref{eq:dx_w_ratio}) to guide the eye.}
	\label{fig:shapeinvariance}
\end{figure}

That means, to maintain the same ratio between scanning amplitude $\Delta x(z)$ and spot radius~$w(z)$ both the translational deflection $X$ as well as the deflection angle $\theta$ have to be $\sqrt{\bar{m}}$ times larger than the waist spot radius $w_0$ and the divergence $\theta_0$ of the deflected beam, respectively.
The magnitude of $\sqrt{\bar{m}}$ then simply determines the spatial amplitude of the scanning motion in relation to the beam size, e.g., $\sqrt{\bar{m}}=1$ yields a rather small scanning motion, where the center of the beam periodically moves between -$w$ and $+w$.
The shape-invariance resulting from the balance between translational and rotational motion is also seen in the calculations in Fig.~\ref{fig:shapeinvariance}a and within an exemplary series of measurements in Fig.~\ref{fig:shapeinvariance}b. 

In the measurements a beam of radius $w_0=\qty{0.3}{\mm}$ was periodically deflected from a translation-rotation mirror pair with a fixed translational amplitude $X$ but various angular deflection amplitudes $\theta$ (see Fig.~2 in the main document for the experimental setup).
The scanning motions were observed to be shape-invariant when translational ($X/w_0$) and rotational displacement ($\theta/\theta_0$) were balanced (green curve in Fig.~\ref{fig:shapeinvariance}b), in accordance with Eq.~(\ref{eq:shapeinvariance}).

Closer inspection of Eq.~(\ref{eq:dx_w_ratio}) reveals that -- as expected -- the beam displacement is determined by translation close to the origin, while it is dominated by rotation in the far field, which can be seen when evaluating the limits $z=0$ and $z\to\infty$: 
\begin{equation}
	\frac{\Delta x(z=0)}{w(z=0)} = \frac{X}{w_0}, \quad \lim_{z\to \infty} \frac{\Delta x(z)}{w(z)} = \frac{\theta}{\theta_0}.
\end{equation}
Hence, in the far field the scanning motion is approximately shape-invariant with the normalized displacement $\Delta x/w$ approaching a constant value, which is solely determined by the rotational/angular parameters (see the set of almost parallel lines for greater values of $z/z_\text{R}$ with their respective limits $\theta/\theta_0$ in Fig.~\ref{fig:shapeinvariance}b).
However, true shape-invariance, where the amplitude of the scanning motion relative to the spot size does not change upon propagation and $\Delta x/w$ is therefore truly independent from $z$, is only achieved by combined translational and rotational displacement.

\bibliography{bib_supplement}